\documentclass[letterpaper,12pt]{article}
\usepackage{bm}
\usepackage{tabularx} 
\usepackage{amsmath}  
\usepackage{graphicx} 
\usepackage[margin=1in,letterpaper]{geometry} 
\usepackage{cite} 
\usepackage[final]{hyperref} 
\usepackage{feynmf}
\hypersetup{
	colorlinks=true,       
	linkcolor=blue,        
	citecolor=blue,        
	filecolor=magenta,     
	urlcolor=blue         
}

\begin{document}

\title{Two dynamical generated $a_0$ resonances by interactions between vector mesons}
\author{Zheng-Li Wang$^{1,2,}$\footnote{Email address:
  \texttt{wangzhengli@itp.ac.cn} }~   and
  Bing-Song Zou$^{1,2,3}$\footnote{Email address:
  \texttt{zoubs@itp.ac.cn} }
  \\[2mm]
  {\it\small$^1$CAS Key Laboratory of Theoretical Physics, Institute
  of Theoretical Physics,}\\
  {\it\small  Chinese Academy of Sciences, Beijing 100190,China}\\
  {\it\small$^2$School of Physics, University of Chinese Academy of Sciences (UCAS), Beijing 100049, China} \\
  {\it\small$^3$School of Physics and Electronics, Central South University, Changsha 410083, China} \\
}
\date{\today}
\maketitle

\begin{abstract}
We study dynamically generated $a_0$ resonances by interactions between vector mesons including their coupling to channels of pseudoscalar mesons within coupled-channel approach. Both vector and pseudoscalar mesons are considered as $t$-channel exchanged mesons for calculating interactions between vector mesons.  Analogous to $a_0(980)$ as a $K\bar K-\pi\eta$ dynamically generated state, there is an $a_0(1710)$ as a coupled channel dynamically generated state near $K^*\bar{K}^*$ threshold.  The channels involved are $\rho\phi,K^*\bar{K}^*,\rho\omega, K\bar{K},\pi\eta$. This $a_0$ mainly decays to $\rho\omega$, $\pi\eta$ and $K\bar{K}$. This pole is much tied to the coupled channel effect. If we turn off either $\rho\phi$ or $K^*\bar{K}^*$, the pole disappears. In addition, it is found that $a_0(1450)$ may also be dynamically generated by $\rho\omega$ interactions due to $\pi$ and $\eta$ exchange. 
\end{abstract}

\section{Introduction}
More and more hadron resonances have been proposed to be hadronic molecules~\cite{Guo:2017jvc} with much more predicted ones to be searched for~\cite{Dong:2021juy,Dong:2021bvy}. Among various approaches for studying hadronic molecules, a quite popular one is the unitary extension of chiral perturbation theory, which has been successfully to study the meson-baryon and meson-meson interactions
at low energy~\cite{Oller:1999ag,Oller:2000fj,Dobado:1996ps,Oller:1998zr,
Oller:1998hw,Oller:1997ti,Oller:2000ma,Molina:2020hde}. 
A well-known example is the $\Lambda(1405)$~\cite{Jido:2003cb}, 
which can be dynamically
generated in the vicinity of the $\pi\Sigma$ and $K^-p$ thresholds. The another
example is $f_0(980)$~\cite{Oller:1997ti,Janssen:1994wn}, 
which is considered to arise due to $\pi\pi$ and $K\bar{K}$ coupled channel
interaction. Some recent works~\cite{Geng:2008gx,Du:2018gyn}
studied the interaction of the nonet of vector mesons themselves
and found a pole with quantum number $I^G(J^{PC})=1^-(0^{++})$ mainly coupling to $\bar{K}^*K^*$ channel. No such $a_0$ around $\bar{K}^*K^*$ threshold is listed in PDG~\cite{PDG}. However, recently both Babar Collaboration~\cite{Lees:2018qrk} and BESII Collaboration~\cite{BESIII:2021anf} have reported strong evidence for a new $a_0(1710)$ resonance with mass nearly degenerate with $f_0(1710)$. The claimed mass is smaller than predicted ones of Refs.\cite{Geng:2008gx,Du:2018gyn} which have ignored the coupled channels of pseudoscalar mesons. 
In this paper, we extend the previous study~\cite{Du:2018gyn} of this resonance by including its coupling to channels of pseudoscalar mesons in addition to vector mesons to see how these more coupled channels influence the pole and result in corresponding partial decay widths to these channels.  For the $t$-channel meson exchanges, besides vector mesons considered in \cite{Geng:2008gx,Du:2018gyn}, we also include pseudoscalar mesons. In addition to the dynamically generated $a_0(1710)$ close to the $\bar{K}^*K^*$ threshold, it is found that $a_0(1450)$ may also be dynamically generated by $\rho\omega$ interactions due to $\pi$ and $\eta$ exchange. 

In the following, we first outline the formalism to the coupled-channel interaction
\cite{Oller:2019opk} in Sect.~\ref{sec:Formalism}, then
in Sect.~\ref{sec:Results}, we give our numerical results and discussion, with a brief summary at the end.

\section{Formalism}\label{sec:Formalism}
\begin{figure}[htp]
  \centering
  \includegraphics[scale=1]{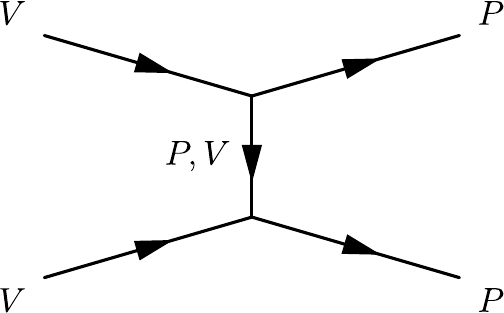}
  \qquad
  \includegraphics[scale=1]{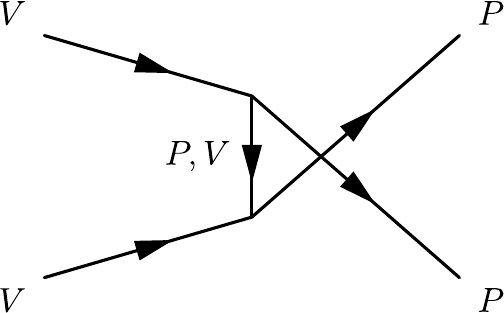}
  \caption{The $t$- and $u$-channel Feynman diagrams} \label{fig:feyndiag}
\end{figure}

The interaction Lagrangian among vector mesons and pseudoscalar mesons is given by
Refs.\cite{Bando:1984ej,Bando:1987br} as the following
\begin{gather} \label{eq:lag}
  \mathcal{L}_{VPP} = -ig \langle V^\mu [P,\partial_\mu P] \rangle, \\
  \mathcal{L}_{V V P}=\frac{G^{\prime}}{\sqrt{2}} \epsilon^{\mu \nu \alpha \beta}\left\langle\partial_{\mu} V_{\nu} \partial_{\alpha} V_{\beta} P\right\rangle.
\end{gather}
with
\begin{equation}
  G'=\frac{3g'^2}{4\pi^2f_\pi} \qquad g'=-\frac{G_VM_V}{\sqrt{2}f^2_\pi},
\end{equation}
where the symbol $\langle \ldots \rangle$ stands for the trace in the $SU(3)$ space
and the coupling constant $g=M_V/2f_\pi$ with $M_V=845.66MeV$ the $SU(3)$-averaged
vector-meson mass, $G_V\simeq 55MeV$ and $f_\pi = 93MeV$ the pion decay constant.
The vector field $V^\mu$ is
\begin{equation}
  V^\mu = \left(
  \begin{array}{ccc}
    \frac{1}{\sqrt{2}} \rho^0 + \frac{1}{\sqrt{2}} \omega & \rho^+ & K^{*+} \\
    \rho^- & -\frac{1}{\sqrt{2}} \rho^0 + \frac{1}{\sqrt{2}} \omega & K^{*0} \\
    K^{*-} & \bar{K}^{*0} & \phi \\
  \end{array} \right)^\mu,
\end{equation}
and the pseudoscalar field $P$ is
\begin{equation}
  P = \left(
  \begin{array}{ccc}
    \frac{1}{\sqrt{2}} \pi^0 + \frac{1}{\sqrt{6}} \eta & \pi^+ & K^+ \\
    \pi^- & -\frac{1}{\sqrt{2}} \pi^0 + \frac{1}{\sqrt{6}} \eta & K^0 \\
    K^- & \bar{K}^0 & -\frac{2}{\sqrt{6}} \eta \\
  \end{array} \right).
\end{equation}
With the Lagrangian given in Eq.~\eqref{eq:lag}, we are able to calculate the
vector-vector to pseudoscalar-pseudoscalar scattering amplitudes. The Feynman diagrams
needed are shown in Fig.~\ref{fig:feyndiag}, where $V$ means vector meson and $P$ means 
pseudoscalar meson.

The amplitudes with isospin-$1$ for the processes
$V(p_1)V(p_2) \to P(p_3)P(p_4)$ are listed in Table~\ref{tab:potential}. 
\begin{table}[htp]
  \centering
  \[ \begin{array}{cc}
    \hline
    \text{Channel} & T^{(1)} \\
    \hline
    \rho\phi \xrightarrow{K,K^*} K\bar{K} & (VP_t^{K}+VP_u^{K})-(VV_t^{K^*}+VV_u^{K^*}) \\
    K^*\bar{K}^* \xrightarrow{K,K^*} \pi\eta & -\frac{\sqrt{6}}{2}(VP_t^{K}+VP_u^{K})+\frac{1}{\sqrt{6}}(VV_t^{K^*}+VV_u^{K^*})\\
    K^*\bar{K}^* \xrightarrow{\pi,\eta,\rho,\omega,\phi} K\bar{K} & (-\frac{1}{2}VP_t^{\pi}+ \frac{3}{2}VP_t^\eta) + (-\frac{1}{2}VV_t^\rho + \frac{1}{2}VV_t^\omega + VV_t^\phi) \\
    \rho\omega \xrightarrow{\rho,\omega} \pi\eta & \frac{2}{\sqrt{3}}(VV_t^{\rho}+VV_t^{\omega})\\
    \rho\omega \xrightarrow{K,K^*} K\bar{K} & -\frac{1}{\sqrt{2}}(VP_t^{K}+VP_u^{K})-\frac{1}{\sqrt{2}}(VV_t^{K^*}+VV_u^{K^*})\\
    \hline
  \end{array} \]
  \caption{The potential of each channel with isospin-$1$} \label{tab:potential}
\end{table}
The convention used to relate the particle basis to the isospin basis is
\begin{equation} \begin{split}
  |\pi^+ \rangle &= -|1,1\rangle \qquad |K^+ \rangle = -|\frac12,\frac12 \rangle, \\
  |\rho^+ \rangle &= -|1,1\rangle \qquad |K^{*+} \rangle = -|\frac12,\frac12 \rangle.
\end{split} \end{equation}
The $VP_{t(u)}$ and $VV_{t(u)}$ correspond to the $t(u)$-channel diagrams with pseudoscalar meson and vector meson exchange, respectively. The superscript
is the particle exchanged. Here, $t=(p_1-p_3)^2$ and $u=(p_1-p_4)^2$ are the usual
Mandelstam variables. The potential is given by
\begin{gather}
  VP_{t(u)}^{ex} =\frac{-4g^2}{t(u)-m^2_{ex}} \epsilon_1 \cdot p_3 \epsilon_2 \cdot p_4, \\
  VV_{t(u)}^{ex} = -\frac{G'^2}{2}\frac{1}{t(u)-m^2_{ex}}\varepsilon_{\mu\nu\alpha\beta} p_1^\mu \epsilon_1^\nu q^\alpha \varepsilon_{\lambda\tau\gamma\delta}p_2^\lambda \epsilon_2^\tau q^\gamma \left(-g^{\beta\delta}+\frac{q^\beta q^\delta}{q^2}\right).
\end{gather}
where the $\epsilon_i$ is the $i$-th polarization vector of the incoming vector meson.
The polarization vector can be characterized by its three-momentum $\mathbf{p}_i$ and
the third component of the spin in its rest frame, and the explicit expression of the
polarization vectors can be found in Appendix A of Ref.~\cite{Gulmez:2016scm}. At the threshold, $s_{th}=(m_1+m_2)^2$, the potential is
\begin{equation}
  VP_t^{ex}=-VV_t^{ex} = \frac{4p^2_f}{\sqrt{3}(m_1m_2-(m_2m_3^2+m_1m_4^2)/\sqrt{s_{th}} +m_{ex}^2)}g^2
\end{equation}
where $p_f$ is the on-shell three momentum of the final state.

In term of these amplitudes with isospin-$1$, we can get the $S$-wave potential via
\cite{Gulmez:2016scm}
\begin{equation} \label{eq:projection} \begin{split}
  T^{(JI)}_{\ell S;\bar{\ell} \bar{S}}(s) &=
  \frac{Y^0_{\bar{\ell}}(\hat{\mathbf{z}})}{\sqrt{2}^N (2J+1)}
  \sum_{\begin{subarray}{c}
    \sigma_1,\sigma_2,\bar{\sigma}_1 \\
    \bar{\sigma}_2,m
  \end{subarray}}
  \int \mathrm{d} \hat{\mathbf{p}}'' Y^m_\ell (\mathbf{p}'')^*
  (\sigma_1 \sigma_2 M | s_1 s_2 S) \\
  &\times (mM\bar{M} | \ell SJ) 
  (\bar{\sigma}_1 \bar{\sigma}_2 \bar{M} | \bar{s}_1 \bar{s}_2 \bar{S})
  (0\bar{M} \bar{M} | \bar{\ell} \bar{S} J) \\
  &\times T^{(I)}(p_1,p_2,p_3,p_4;\epsilon_1,\epsilon_2,\epsilon_3,\epsilon_4).
\end{split} \end{equation}
with $s=(p_1+p_2)^2$ the usual Mandelstam variable, $M=\sigma_1+\sigma_2$ and
$\bar{M} = \bar{\sigma}_1 + \bar{\sigma}_2$. And $N$ accounts for the identical
particles, for example
\begin{flalign}
  N &= 2 \text{ for } \rho\rho \to \pi\pi, & \\
  N &= 1 \text{ for } \rho\rho \to K\bar{K}, & \\
  N &= 0 \text{ for } \omega\phi \to K\bar{K}. &
\end{flalign}
Like vector scattering $ VV \to V V$, the partial wave projection
Eq.~\eqref{eq:projection} for a $t$-channel exchange amplitude of
$V V \to P P$ would also develop a left-hand cut via~\cite{Du:2018gyn}
\begin{equation} \begin{split}
  \frac12 & \int_{-1}^{+1} \mathrm{d} \cos \theta \frac{1}{t-m^2_{ex}+i\epsilon} =
  -\frac{s}{\sqrt{\lambda (s,m_1^2,m_2^2) \lambda (s,m_3^2,m_4^2)}} \\
  &\times \log \frac{m_1^2 + m_2^2 - \frac{(s+m_1^2-m_2^2)(s+m_3^2-m_4^2)}{2s} -
  \frac{\sqrt{\lambda (s,m_1^2,m_2^2) \lambda (s,m_3^2,m_4^2)}}{2s} -m_{ex}^2 +i\epsilon}
  {m_1^2 + m_2^2 - \frac{(s+m_1^2-m_2^2)(s+m_3^2-m_4^2)}{2s} +
  \frac{\sqrt{\lambda (s,m_1^2,m_2^2) \lambda (s,m_3^2,m_4^2)}}{2s} -m_{ex}^2 +i\epsilon}.
\end{split} \end{equation}
with $\lambda (a,b,c) = a^2+b^2+c^2-2ab-2bc-2ac$ the K\"all\'en function. In vector
scattering $V V \to V V$, left-hand cuts are smoothed by the $N/D$ method~\cite{Chew:1960iv,Bjorken:1960zz}. 
As for the scattering $V V \to P P$,
all left-hand cuts are located below the $PP$ threshold, which are far away from
the energy region we are interested in, so we do not deal with these cuts.

The basic equation to obtain the unitarized $T$-matrix is
\begin{equation}
  T^{(JI)} (s) = \left[ 1-V^{(JI)}(s) \cdot G(s) \right]^{-1} \cdot V^{(JI)}(s).
\end{equation}
Here $V^{(JI)}$ denotes the partial-wave amplitudes and $G(s)$ is a diagonal matrix
made up by the two-point loop function $g_i(s)$,
\begin{equation}
  g_i(s) = i \int \frac{\mathrm{d}^4 q}{(2\pi)^4} \frac{1}
  {(q^2-m^2_{i1}+i\epsilon) ((P-q)^2 -m^2_{i2} +i\epsilon)}.
\end{equation}
with $P^2=s$ and $m_{i1,2}$ the masses of the particles in the $i$-th channel. The pole
position is at the zeros of determinant
\begin{equation}
  \text{Det} \equiv \text{det} \left[ 1- V^{(JI)}(s) \cdot G(s) \right].
\end{equation}
The above loop function is logarithmically divergent and can be calculated with a
once-subtracted dispersion relation or using a regularization $f_\Lambda (q)$
\begin{equation}
  g_i(s) = i \int \frac{\mathrm{d}^4 q}{(2\pi)^4} \frac{f^2_\Lambda (q)}
  {(q^2-m^2_{i1}+i\epsilon) ((P-q)^2 -m^2_{i2} +i\epsilon)}.
\end{equation}
after the $q^0$ integration is performed by choosing the contour in the lower half of
the complex plane, we get
\begin{equation}
  g_i(s) = \int_0^\infty \frac{|\mathbf{q}|^2 \mathrm{d}|\mathbf{q}|}{(2\pi)^2}
  \frac{\omega_{i1} + \omega_{i2}}{\omega_{i1}\omega_{i2}
  (s-(\omega_{i1} + \omega_{i2})^2 + \mathrm{i}\epsilon)} f_\Lambda^2(|\mathbf{q}|).
\end{equation}
where $\mathbf{q}$ is the three-momentum and
$\omega_{i1,2} = \sqrt{\mathbf{q}^2 + m^2_{i1,2}}$. In order to proceed we need
to determine $f_\Lambda (\mathbf{q})$. There are two kinds of choices, sharp cutoff
and smooth cutoff, typically:
\begin{equation}
  f_\Lambda (\mathbf{q}) = \left\{
    \begin{array}{l}
      \Theta(\Lambda^2 - \mathbf{q}^2) \\
      \exp\left[-\frac{\mathbf{q}^2}{\Lambda^2}\right]\\
    \end{array} \right.
\end{equation}

In order to compare with the previous results of coupled channel approach~\cite{Du:2018gyn}, the same sharp cutoff is used in this paper when channels with pseudoscalar mesons are included in addition. To explore the
position of the poles we need to take into account
the analytical structure of these amplitudes in the different Riemann sheets. By
denoting $q_{on}$ for the CM tri-momentum of the particles $1$ and $2$ in the 
$i$-th channel
\begin{equation}
  q_i^\text{on} = \frac{\sqrt{(s-(m_{i1}-m_{i2})^2) (s-(m_{i1}+m_{i2})^2)}}{2\sqrt{s}}.
\end{equation}
As the quantity is two-valued itself~\cite{Doring:2009yv}, 
we need to distinguish the two Riemann sheets
of $q_i^\text{on}$ uniquely according to
\begin{equation}
  q_i^{\text{on}>} = \left\{
    \begin{array}{cl}
      -q_i^\text{on} & \text{if } \text{Im} q_i^\text{on} < 0 \\
      q_i^\text{on} & \text{else} \\
    \end{array}\right.
\end{equation}
And the analytic continuation to the second Riemann sheet is  given by 
\begin{equation}
  g^{(2)}_i (s) = g_i (s) + \frac{i}{4\pi} \frac{q_i^{\text{on}>}}{\sqrt{s}}.
\end{equation}

\section{Numerical results and discussion }\label{sec:Results}

Firstly, within the isospin formalism the $\rho\rho$ states obey Bose-Einstein statistics, so that only states with even $l+S+I$ are allowed, which means that in our case $\rho\rho$ channel is ruled out.

For $K^*\bar{K}^*$ single channel, we already know that there is a bound state for $I=0$ sector, and the potential is about $V_{11}^{(0)}\simeq -7.7g^2$~\cite{Du:2018gyn}. We label the $K^*\bar{K}^*$ as channel $1$, and 
the remain channel indices are listed in Table~\ref{tab:index}. While for the $I=1$ sector, the potential is about $V_{11}^{(1)}\simeq -0.98g^2$, which is too weak to produce a bound state.

\begin{table}
  \centering
  \[\begin{array}{ccc}
    \hline
    \text{Channel index} & \text{Channel} & \text{Threshold}~(GeV) \\
    \hline
    1 & K^*\bar{K}^* & 1.784 \\
    2 & \rho\omega & 1.552 \\
    3 & \rho\phi & 1.79 \\
    4 & K\bar{K} & 0.99 \\
    5 & \pi\eta & 0.686 \\
    \hline
  \end{array}\]
  \caption{Channel indices and threshold energies}\label{tab:index}
\end{table}

Then we turn on additional channels to study their influence on the mass and width of the resonance. The
$T$-matrix for a single channel $a$ is given by 
\begin{equation}
  T_{aa} = \frac{V_{aa}}{1-V_{aa}g_a}.
\end{equation}

If we turn on another channel $b$, then the $T$-matrix for $a\to a$ becomes to be
\begin{equation}
  T_{aa} = \frac{V_{aa}+\frac{V_{ab}^2g_b}{1-V_{bb}g_b}}
  {1-(V_{aa}+\frac{V_{ab}^2g_b}{1-V_{bb}g_b})g_a}.
\end{equation}

Compared to the single channel, $V_{aa}$ is replaced by
\begin{equation}
  V_{eff} = V_{aa}+\frac{V_{ab}^2g_b}{1-V_{bb}g_b}.
\end{equation}

Denoting the second term as
\begin{equation}
  V' = \frac{V_{ab}^2g_b}{1-V_{bb}g_b},
\end{equation}
then $T_{aa}$ can be written as
\begin{equation}
  T_{aa} = \frac{V_{aa}+V'}{1-(V_{aa}+V')g_a}
  = \frac{(1+\alpha)V_{aa}}{1-(1+\alpha)V_{aa}g_a}.
\end{equation}
with $\alpha = V'/V_{aa}$. For calculating the loop integral $g_a$ for channels of vector mesons, we use the same sharp cutoff as in the previous study~\cite{Du:2018gyn}. However, if we turn on the channels of pseudo-scalar mesons, the t-channel exchanged pseudo-scalar meson in $VV \to PP$ is mostly off shell for the calculation of $V'$, implementation of some off-shell form factors is necessary~\cite{Molina:2008jw}. Adjustment of the sharp cutoff parameter $q_{max}$ for the loop integration has no influence for the imaginary part of $g_b$, which corresponds to two on-shell pseudo-scalar mesons with an off-shell t-channel exchanged pseudo-scalar meson. To take into account this off-shell effect of the t-channel exchanged meson, the same kind of monopole form factor as in our triangle loop approach as well as in Ref.\cite{Molina:2008jw}     
\begin{equation}
  F = \frac{\Lambda^2 - m^2_{ex}}{\Lambda^2-q^2}
\end{equation}
is included at each $VPP$ vertex for the exchanged pseudo-scalar meson with momentum $q$. For the $VV \to PP$ process, the t-channel exchange meson is much more off-shell than corresponding $VV \to VV$ case. 

Firstly, we turn on the $\rho\phi$ channel, it has a contribution about $V_{13}^2g_3(s) \simeq -4g^2$ near the $K^*\bar{K}^*$ threshold, together with $V_{11} \simeq -0.98g^2$, $V_{eff} \simeq -5g^2$ and there will be a bound state below the $K^*\bar{K}^*$ threshold. We find that the coupled $K^*\bar{K}^*$ and $\rho\phi$ channels are necessary to produce the bound state. There would be no bound state near the $K^*\bar{K}^*$ threshold for ether $K^*\bar{K}^*,\rho\omega,K\bar{K},\pi\eta$ coupled system or $\rho\phi,\rho\omega,K\bar{K},\pi\eta$ coupled system .

Then we turn on other channel one by one. For $\rho\phi-K^*\bar{K}^*-\rho\omega$ system, the pole position is about $1.74-i0.03(GeV)$. For the the $\rho\phi-K^*\bar{K}^*-K\bar{K}$ system, the pole position is about $1.78-i0.07(GeV)$. For the $\rho\phi-K^*\bar{K}^*-\pi\eta$ system, the pole position is about $1.78-i0.02(GeV)$. 

Finally, we turn on all channels
\begin{figure}
  \centering
  \includegraphics[scale=0.5]{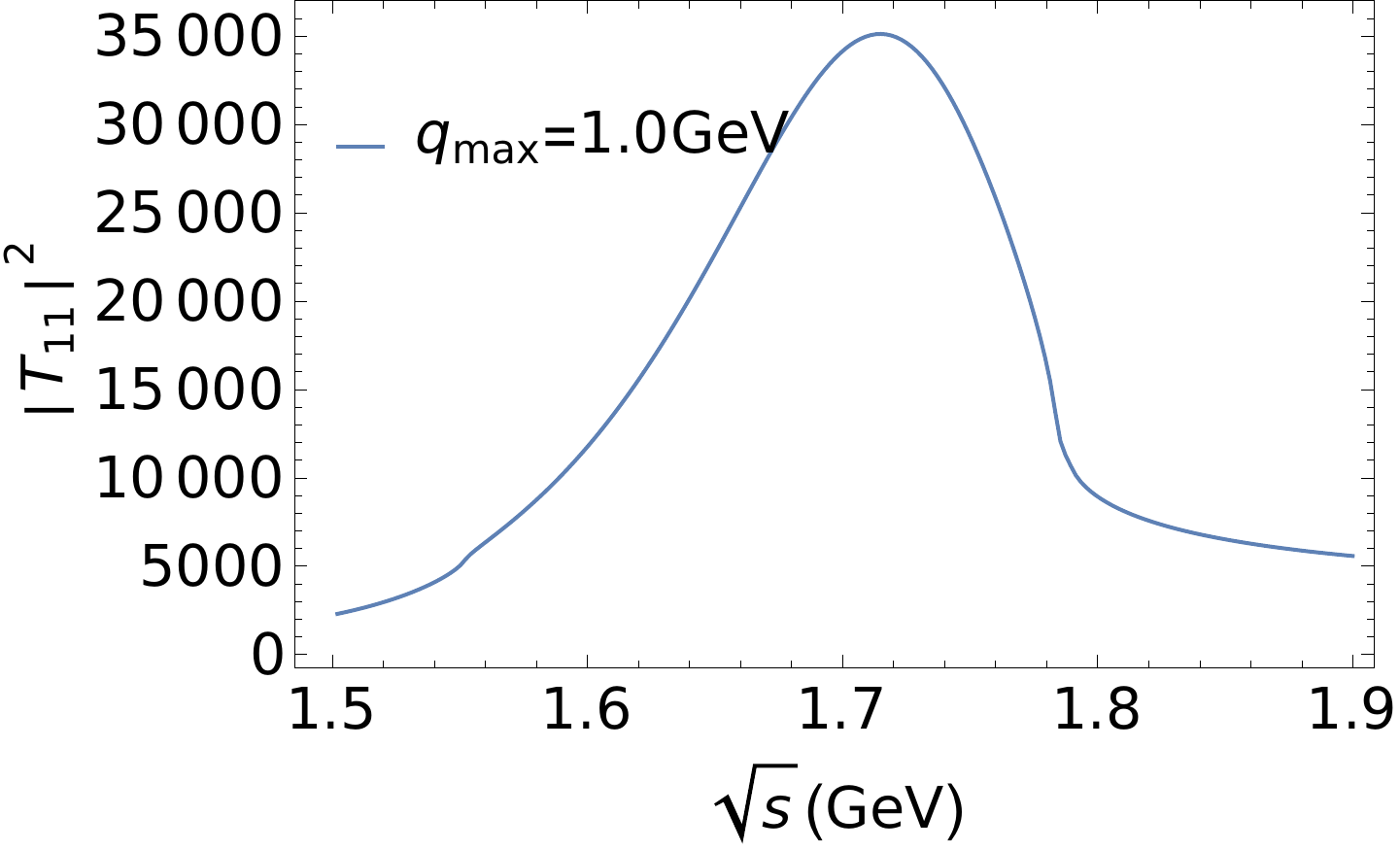}
  \includegraphics[scale=0.5]{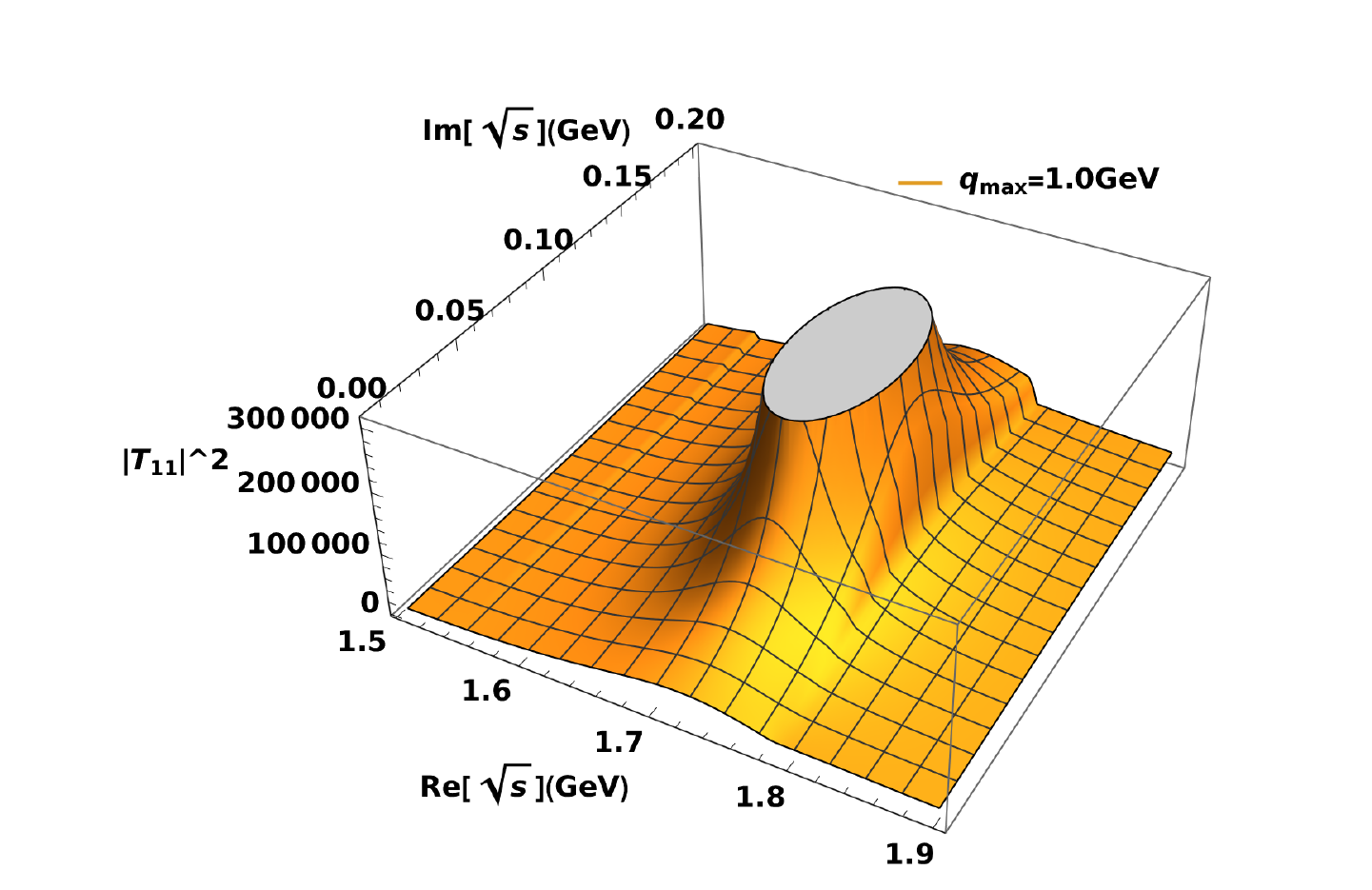}
  \caption{$|T_{11}|^2$ for $\Lambda=1.0 GeV$ and $q_{max}=1.0GeV$ in 2-dim and 3-dim}\label{fig:t11all}
\end{figure}
and show the $|T_{11}|^2$ in 2-dim and 3-dim with $q_{max}=1.0 GeV$ in Fig.~\ref{fig:t11all}. The pole position
of the resonance is listed in Table~\ref{tab:pole} for different cutoffs.
\begin{table}
  \centering
  \[\begin{array}{cccccc}
    \hline
    q_{max}(GeV) & 0.9 & 1.0 & 1.1 \\
    \hline
    Pole(GeV) & 1.76-0.09i & 1.72-0.10i & 1.69-0.11i \\
    \hline
  \end{array}\]
  \caption{The resonance pole for different cutoffs}\label{tab:pole}
\end{table}
We find that the real part of the resonance is about $1720 MeV$ and the width is about
$200 MeV$ with $q_{max}=1.0GeV$. 
With the effective coupling constant of $a_0\bar K^*K^*$ determined by its binding energy, the partial decay width of 
$a_0\to\bar K^*K^*\to K\pi\bar K\pi$ can be calculated straightforwardly to be around 21 MeV, together with the two body decays, the total decay width of $a_0$ is about $220MeV$, larger than $f_0(1710)$. Compared with previous results of Refs.\cite{Geng:2008gx,Du:2018gyn} which ignored the couplings of vector mesons to pseudoscalar mesons, the mass of the resonance gets smaller to be nearly degenerate with its isoscalar partner $f_0(1710)$ and fits in the experimental claimed one $a_0(1710)$ very well. 

For the unitary coupled channel approach, we can calculate each partial decay width via~\cite{PDG}
\begin{equation}
  \Gamma_{R\to a} = \frac{|\tilde{g}_a|^2}{M_R}
  \rho_a(M^2_R).
\end{equation}
with $\tilde{g}_a = \mathcal{R}_{ba}/\sqrt{\mathcal{R}_{bb}}$ and
$\rho_a$ the two-body phase space. The residues may be calculated via an integration 
along a closed contour around the pole using
\begin{equation}
  \mathcal{R}_{ba}=-\frac{1}{2\pi i} \oint ds \mathcal{M}_{ba}.
\end{equation}
The partial decay widths obtained this way are~\ref{tab:width}
\begin{table}[h]
  \centering
  \[\begin{array}{cccccc}
    \hline
    \Gamma(K^*\bar{K}^* \to \rho\omega)&\Gamma(K^*\bar{K}^* \to K\bar{K})&\Gamma(K^*\bar{K}^* \to \pi\eta) \\
    \hline
    61.0MeV & 74.4MeV & 66.9MeV \\
    \hline
    \Gamma(\rho\phi \to \rho\omega)&\Gamma(\rho\phi \to K\bar{K})&\Gamma(\rho\phi \to \pi\eta)\\
    \hline
    60.8MeV & 74.2MeV & 66.6MeV\\
    \hline
  \end{array}\]
  \caption{The partial decay widths}\label{tab:width}
\end{table}

Note that there is no tree diagram for $\rho\phi \to \rho\omega$, which means that $V_{34}=0$, but $T_{34}$ is not $0$ due to the strongly coupled channels of $\rho\phi$ and $K^*\bar{K}^*$ with nearly degenerate thresholds. In both cases the total widths are about $200MeV$.
The relative ratio of the resonance decaying to $\rho\omega,K\bar{K},\pi\eta$ is about $1:1:1$. While in Ref.~\cite{Oset:2012zza}, the ratio is about $2:1:1$, since for $VV \to PP$ we include also vector meson exchange in addition to pseudoscalar meson exchange considered in Ref.~\cite{Oset:2012zza}, the decay widths to pseudoscalar channel become larger. The residues of channel $K^*\bar{K}^*$ is about $8731-i2200(MeV)$ and $\rho\phi$ is about $6013-i1785(MeV)$, which means that the coupling of the resonance to $\rho\phi$ cannot be neglected. Indeed, if we turn off the $\rho\phi$, the resonance disappears.



Up to now, for the $I=0$ sector, there are three hadronic molecules calimed to be dynamically generated from the $K\bar{K}, \rho\rho, K^*\bar{K}^*$ interaction through vector meson exchange potentials, which are assigned to $f_0(980),f_0(1500),f_0(1710)$, respectively~\cite{Oller:1997ti,Wang:2019niy,Wang:2021jub}. For the $I=1$ sector, the attractive potentials for the $K\bar{K}$ and $K^*\bar{K}^*$ channels are much weaker than the corresponding isoscalar cases. No bound states can be formed  for each single channel.  The corresponding $a_0(980)$ and $a_0(1710)$ can only be dynamically generated through strong coupled channel effects.  The $a_0(980)$ arises due to the interaction in the $K\bar{K}-\pi\eta$ system, while $a_0(1710)$ close to $K^*\bar{K}^*$ threshold arises due to interaction in the $\rho\phi-K^*\bar{K}^*-\rho\omega-K\bar{K}-\pi\eta$ system. Then corresponding to the $f_0(1500)$ as the isoscalar $\rho\rho$ bound state,  we also examine whether $a_0(1450)$ can be dynamically generated by the $\rho\omega-K\bar{K}-\pi\eta$ interaction.

\begin{figure}[h]
  \centering
  \includegraphics[scale=0.5]{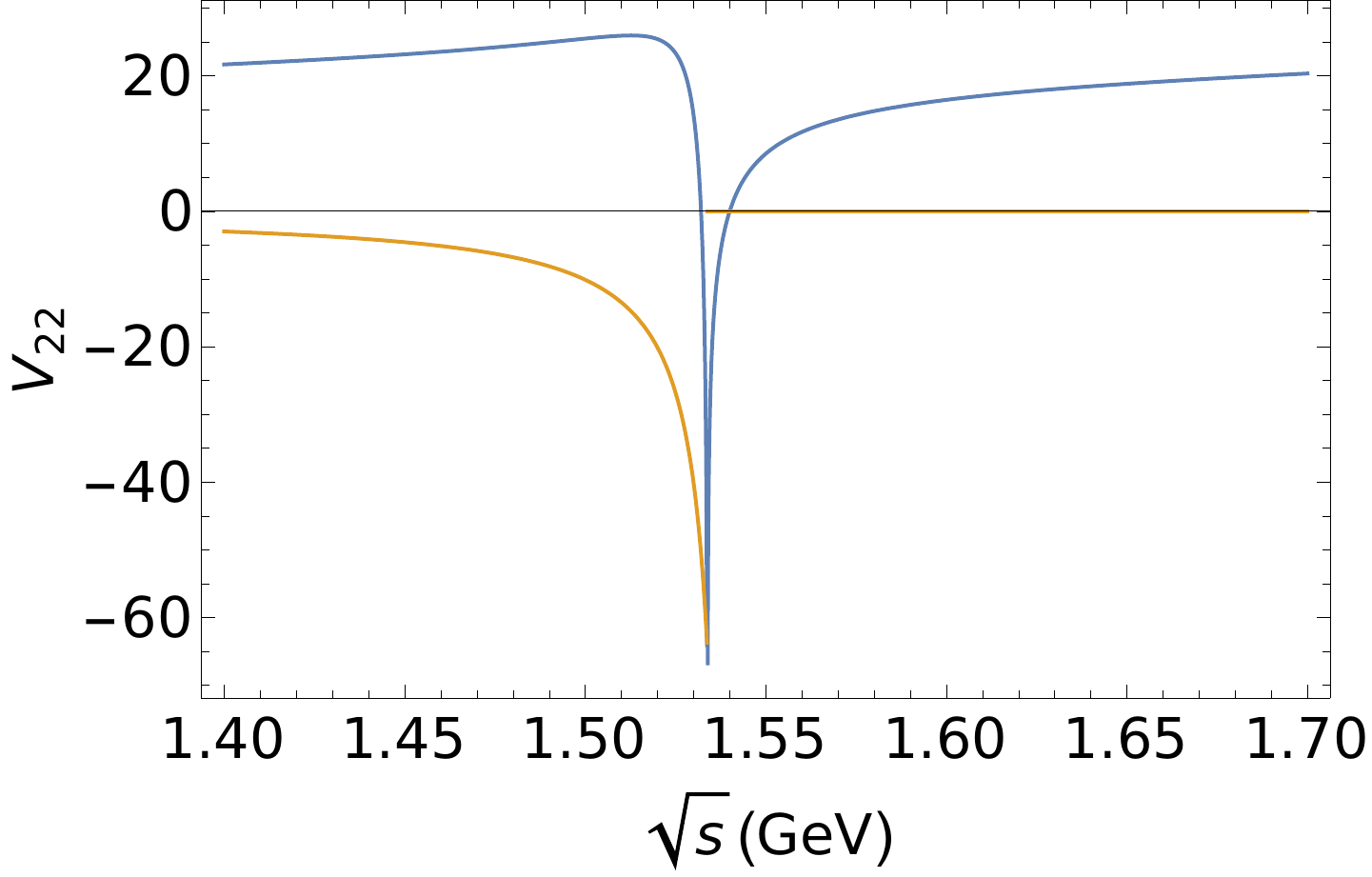}
  \includegraphics[scale=0.5]{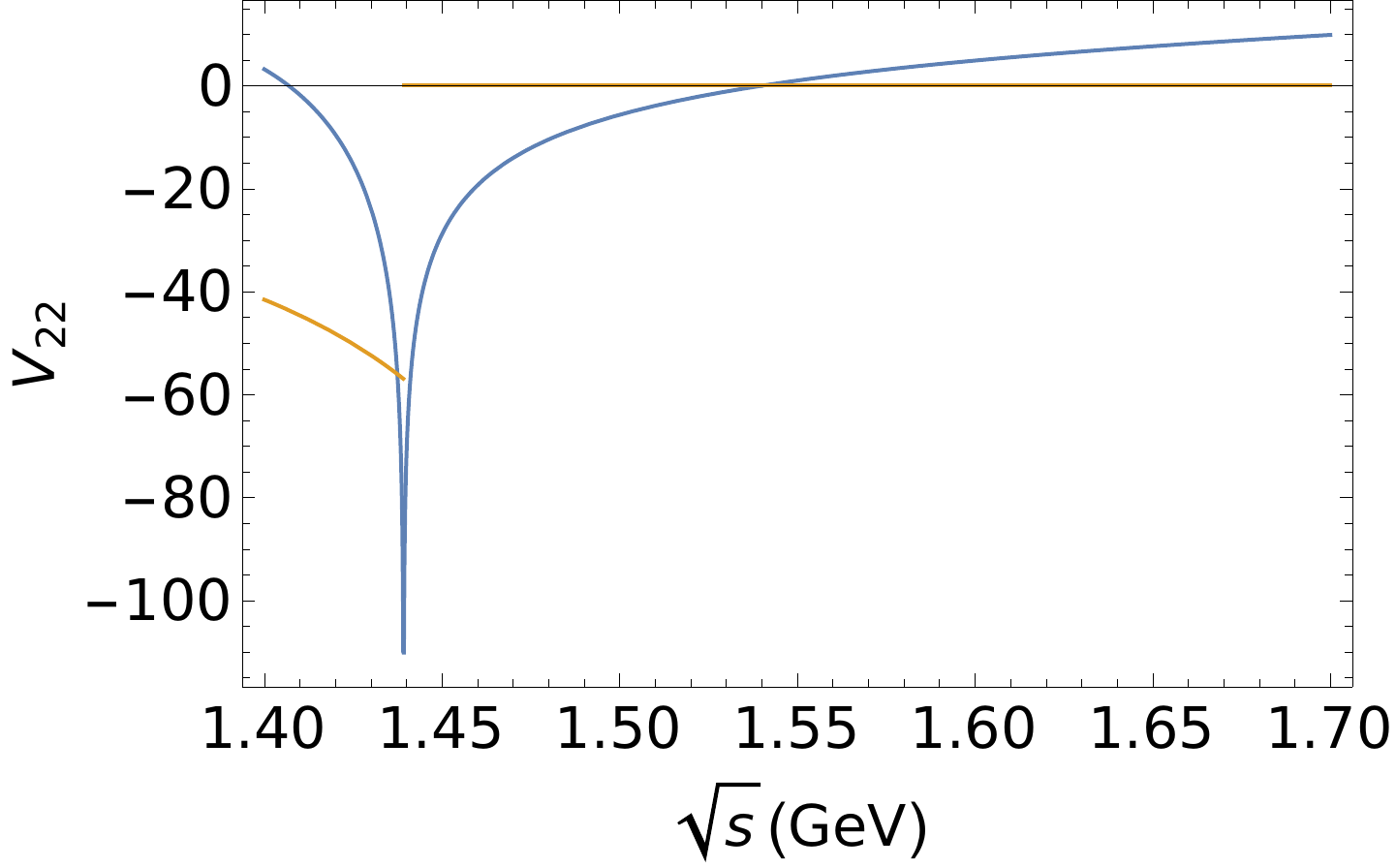}
  \caption{Potential of $\rho\omega \to \rho\omega$, the left one is $\pi$ exchange and the right one is $\eta$ exchange. The blue line is real part of the potential and the yellow line is imaginary part}\label{fig:rhoomega}
\end{figure}

For the $\rho\omega$ interaction, no $t$-channel vector meson exchange is allowed. Therefore no dynamically generated $a_0$ is found around $\rho\omega$ threshold in the previous studies~\cite{Geng:2008gx,Du:2018gyn}. However, when taking into account the pseudoscalar meson exchange force,  the potential of $\rho\omega \xrightarrow{\pi,\eta} \rho\omega$ is shown in Fig.~\ref{fig:rhoomega}

\begin{figure}[h]
  \centering
  \includegraphics[scale=1.5]{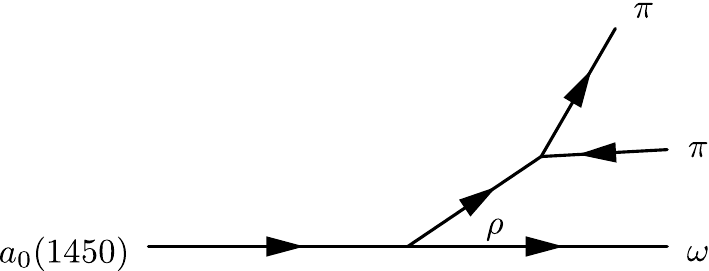}
  \caption{Three body decay $a_0(1450) \to \omega\pi\pi$}\label{fig:decay}
\end{figure}

The potential is $0$ at the threshold and the left hand cut is close to the threshold. The $\pi$ exchange is very steep at the threshold, so when the $N/D$ method is used, the potential is strong enough to produce a pole below the $\rho\omega$ threshold. However, $\eta$ exchange is very weak. When scanning on the complex plane, we find the pole position is about $1.5-i0.015(GeV)$, and $\Gamma(a_0(1450) \to K\bar{K})=13.2MeV$, $\Gamma(a_0(1450) \to \pi\eta)=14.7MeV$, which gives a relative ratio of 0.9 in consistent with the PDG value. Then we calculate the three body decay $a_0(1450) \to \omega\pi\pi$ as shown in Fig.\ref{fig:decay}. The three-body decay width is about $100 MeV$, and results in the total decay width to be around 128 MeV, smaller than its PDG average value. But different experiments report quite different values for both its mass and width. Some of them are in fact consistent with our theoretical values.

In summary, we extend the coupled channel interaction of vector mesons by including channels of pseudo-scalar mesons in addition, using the unitary coupled-channel approach in the $I=1$ sector. The pole near the $K^*\bar{K}^*$
threshold remains to be there, but with its mass lower to be nearly degenerate with its isoscalar partner $f_0(1710)$. The mass is consistent with the $a_0(1710)$ newly reported by Babar Collaboration~\cite{Lees:2018qrk} and BESII Collaboration~\cite{BESIII:2021anf}. The ratio of its decays to $\rho\omega,K\bar{K},\pi\eta$ is predicted to be about $1:1:1$.
With pseudo-scalar meson exchange included, another pole is found to be just below $\rho\omega$ threshold, with its mass nearly degenerate with isoscalar $\rho\rho$ bound state $f_0(1500)$ and in consistent with $a_0(1450)$ within experimental uncertainties. Its dominant decay mode is predicted to be $\omega\pi\pi$.  

Previously, both $f_0(1500)$ and $f_0(1710)$ have been considered as glueball candidates and extensively studied within the quarkonia-guleball mixing picture~\cite{Amsler:1995td,Albaladejo:2008qa,Close:1996yc, Close:2005vf, Giacosa:2005zt, Chanowitz:2005du,Chao:2007sk}. With the success of the new possible configuration as hadron molecule to explain their properties~\cite{Wang:2019niy,Wang:2021jub}, to pin down their nature, it is crucial to study also their relevant iso-vector $a_0$ mesons. While scalar glueballs are not ecpected to have iso-vector partners, the $f_0(1500)$ and $f_0(1710)$ as scalar hadronic molecules should have nearly degenerate iso-vector partners, $a_0(1450)$ and $a_0(1710)$, respectively, just as nearly degenerate $f_0(980)$ and $a_0(980)$ as $\bar KK$ molecules. The $a_0(1710)$ is still not listed in PDG~\cite{PDG} and the dominant decay mode of $a_(1450)$ is still not identified. To establish these two iso-vector hadronic molecules, it would be very useful to study $\gamma\gamma\to\eta\pi$, $\bar KK$, $\omega\pi\pi$ at Belle2 experiment.

Recently, the newly observed $J^{PC}=1^{-+}$ hybrid candidate $\eta_1(1855)$~\cite{BESIII:2022riz} is also explained as a hadronic molecule~\cite{Dong:2022cuw}. It is very likely that the unquenching dynamics leading to hadonic molecules would be the dominant excitation mechanism for hadrons and make it impossible to form glueballs or hybrids.

\section*{Acknowledgments}

We thank useful discussions and valuable comments from Feng-Kun Guo, Ulf-G. Mei{\ss}ner and Jia-Jun Wu. 
This work is supported by the NSFC and the Deutsche Forschungsgemeinschaft (DFG, German Research
Foundation) through the funds provided to the Sino-German Collaborative
Research Center TRR110 “Symmetries and the Emergence of Structure in QCD”
(NSFC Grant No. 12070131001, DFG Project-ID 196253076 - TRR 110), by the NSFC 
Grant No.11835015, No.12047503, and by the Chinese Academy of Sciences (CAS) under Grant No.XDB34030000.


\end{document}